\documentclass[11pt,a4paper]{article}

\usepackage[english]{babel}
\usepackage[backend=biber,style=ieee]{biblatex}
\usepackage{booktabs}
\usepackage{epstopdf}
\usepackage{fancyhdr}
\usepackage{float}
\usepackage[top=1in,left=1in,right=1in,bottom=1in]{geometry}
\usepackage{graphicx}
\usepackage[hidelinks]{hyperref}
\usepackage[utf8]{inputenc}
\usepackage{lastpage}
\usepackage{lmodern}
\usepackage{nameref}

\raggedright

\begin{document}

\begin{flushleft}
{\Large
\textbf\newline{\texttt{crate2bib}: Citing Rust crates made easy}
}
\newline
\\
\href{https://orcid.org/0009-0001-0613-7978}{\includegraphics[width=1em]{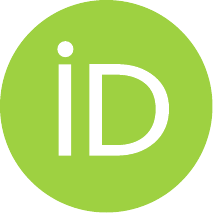}}%
\hspace{1mm}Jonas Pleyer\textsuperscript{1}\\
\bigskip
\bf{1} Freiburg Center for Data Analysis and Modelling and AI (FDMAI), University of Freiburg, Freiburg, Germany\\
\bigskip
* jonas.pleyer@fdm.uni-freiburg.de

\end{flushleft}

\section*{Abstract}
\texttt{crate2bib} is a collection of tools designed to convert Rust crates hosted on
\href{https://crates.io}{crates.io} into bibliography entries.
It queries the server, extracts metadata from the given crate and also searches for possible
\texttt{CITATION.cff} files within the repository that hosts the code of the crate in interest.
From this information, it formats the provided information such as name, version, authors and
generates entries for all available candidates.
With this approach, crates can be cited easily and existing citations for published crates can be
found.
The tool can be used as a webapp, python package, command-line utility or Rust crate.

\begin{figure}[H]
    \centering
    \includegraphics[width=0.6\textwidth]{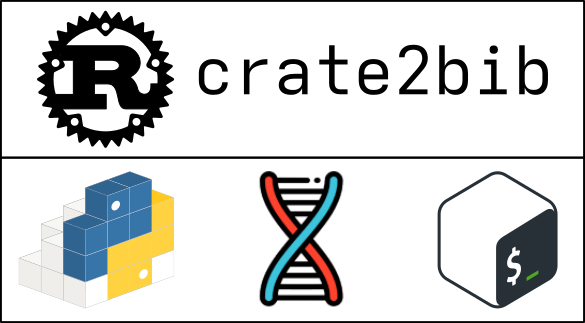}
    \caption{
        The various components of \texttt{crate2bib} are built on a Rust crate which is reused to
        provide a python library, a webapp and a CLI tool.
    }
\end{figure}

\section*{Statement of Need}
\texttt{crate2bib} aims to provide a simple, user-friendly platform to cite Rust software in research
papers, theses or technical documentation.
As the field of scientific computing continues to grow, it's of best scientific practice to
attribute contributions of libraries that underpin novel results and approaches.
Instead of having to manually compile bibliography entries or manually search for the necessary
information, which is prone to errors, \texttt{crate2bib} utilizes the published metadata which was
already provided by the authors of the software.
Furthermore, the provided methods can also be used for automation processes.
The result is saved time as well as concisely structured citations which can be readily used within
scientific publishing workflows.
Incidently, this paper already uses the provided methods extensively.

Over the course of the past years, multiple citation strategies have emerged within the \LaTeX
ecosystem.
The BibTeX format has been the de-facto standard and still remains widely used.
It was extended by natbib \cite{Daly2025} with more citation styles and citing commands.
Another alternative is BibLaTeX which provides a "[..] complete reimplementation of
the bibliographic facilities provided by \LaTeX" \cite{biblatex2025}.
Tools such as \cite{Gerosa2017} filltex target specific areas of research, enabling queries to ADS and
INSPIRE databases.
Within the Github ecosystem, the
\href{https://citation-file-format.github.io/}{\texttt{CITATION.cff}} format
\cite{Druskat2021} was introduced that is being adapted by many maintainers.
It is only natural that conversion tools have emerged, allowing the mapping of both directions.
BibTeX2CFF \cite{Hahn2023} converts from BibTeX to the CFF format, while
cffconvert \cite{Spaaks2021} acts in the reverse direction.
This project also provides tools such as an initializer \cite{Spaaks2023} to quickly create
new \texttt{CITATION.cff} files.

In order to connect the rapidly growing landscape of Rust-based software with academic publishing
standards, it is utmost helpful to have methods available which can generate
bibliography entries and aid in automation processes.

\section*{Library Details}

The \texttt{crate2bib} Rust crate is the foundational component which enables the remaining tools (webapp,
python bindings, cli utility).
It holds the core components and logic and uses the \texttt{reqwest} \cite{McArthur2025} crate
together with \texttt{crates\_io\_api} \cite{Herzog2025} to access information from
\href{https://crates.io}{crates.io} and repository hosting platforms such as
\href{https://github.com}{github.com} and \href{https://codeberg.org/}{codeberg.org}.
The latter crate was forked by the author \cite{Pleyer2025} in order to provide WASM support which was
required for the webapp but previously missing in its original version.
To parse the obtained information, \texttt{crate2bib} relies on biblatex \cite{Haug2025} and citeworks-cff
\cite{Saparelli2022}.
Python bindings are produced with pyo3 \cite{pyo32025} and published with maturin \cite{maturin2025}
while the webapp is built with dioxus \cite{Kelley2025}.

\begin{table}[H]
    \centering
    \begin{tabular}{ll}
        \toprule
        Tool & Link\\
        \midrule
        Webapp & \href{https://jonaspleyer.github.io/crate2bib/}{jonaspleyer.github.io/crate2bib/}\\
        Python Bindings & \href{https://pypi.org/project/crate2bib/}{pypi.org/project/crate2bib/}\\
        CLI tool & \href{https://crates.io/crates/crate2bib-cli}{crates.io/crates/crate2bib-cli}\\
        Rust crate & \href{https://crates.io/crates/crate2bib}{crates.io/crates/crate2bib}\\
        \bottomrule
    \end{tabular}
    \caption{List of provided tools by \texttt{crate2bib}.}
\end{table}

\newpage
\section*{Acknowledgements}
Image credit: Python Software Foundation, Dioxus Labs, Wikimedia Commons

\printbibliography

@article{Gerosa2017,
    title = {filltex: Automatic queries to ADS and INSPIRE databases to fill
             LaTex bibliography},
    volume = {2},
    ISSN = {2475-9066},
    url = {http://dx.doi.org/10.21105/joss.00222},
    DOI = {10.21105/joss.00222},
    number = {13},
    journal = {The Journal of Open Source Software},
    publisher = {The Open Journal},
    author = {Gerosa, Davide and Vallisneri, Michele},
    year = {2017},
    month = may,
    pages = {222},
}

@software{pyo32025,
    author = {},
    title = {{pyo3}: Bindings to Python interpreter},
    url = {https://github.com/pyo3/pyo3},
    date = {2025-10-21},
    version = {0.27.1},
    license = {MIT OR Apache-2.0},
}

@software{maturin2025,
    author = {konsti},
    title = {{maturin}: Build and publish crates with pyo3, cffi and uniffi
             bindings as well as rust binaries as python packages},
    url = {https://github.com/pyo3/maturin},
    date = {2025-10-07},
    version = {1.9.6},
    license = {MIT OR Apache-2.0},
}

@software{Spaaks2021,
    author = {Spaaks, Jurriaan H.},
    license = {Apache-2.0},
    month = sep,
    title = {{cffconvert}},
    url = {https://github.com/citation-file-format/cffconvert},
    version = {3.0.0a0},
    year = {2021},
}

@software{Druskat2021,
    author = {Druskat, Stephan and Spaaks, Jurriaan H. and Chue Hong, Neil and
              Haines, Robert and Baker, James and Bliven, Spencer and Willighagen
              , Egon and Pérez-Suárez, David and Konovalov, Olexandr},
    doi = {10.5281/zenodo.5171937},
    license = {CC-BY-4.0},
    month = aug,
    title = {{Citation File Format}},
    version = {1.2.0},
    year = {2021},
    url = {https://github.com/citation-file-format/citation-file-format},
}

@software{Pleyer2025,
    author = {Jonas Pleyer},
    title = {{crates\_io\_api-wasm-patch}: WASM-compatible patch of crates\_io\_
             api},
    url = {https://github.com/jonaspleyer/crates-io-api},
    date = {2025-01-27},
    version = {0.12.1},
    license = {MIT/Apache-2.0},
}

@software{Herzog2025,
    author = {Christoph Herzog},
    title = {{crates\_io\_api}: API client for crates.io},
    url = {https://github.com/theduke/crates-io-api},
    date = {2025-08-20},
    version = {0.12.0},
    license = {MIT/Apache-2.0},
}

@software{Hahn2023,
    author = {Hahn, Anselm},
    doi = {10.5281/zenodo.8007540},
    month = jun,
    title = {{Convert from bibtex to CITATION.cff}},
    version = {v0.2.1},
    year = {2023},
}

@software{Kelley2025,
    author = {Jonathan Kelley},
    title = {{dioxus}: Build fullstack web, desktop, and mobile apps with a
             single codebase.},
    url = {https://github.com/DioxusLabs/dioxus/},
    date = {2025-11-06},
    version = {0.7.1},
    license = {MIT OR Apache-2.0},
}

@software{biblatex2025,
    author = {Philipp Lehman, Philip Kime, Joseph Wright and Audrey Boruvka},
    title = {BibLaTeX},
    url = {https://github.com/plk/biblatex},
    version = {3.20},
    license = {LaTeX Project Public License},
}

@software{Daly2025,
    author = {Patrick W. Daly},
    title = {Natural Sciences Citations and References},
    version = {8.31b},
    license = {LaTeX Project Public License},
    url = {https://ctan.org/pkg/natbib},
}

@software{Spaaks2023,
    author = {Spaaks, Jurriaan H. and Verhoeven, Stefan and Diblen, Faruk and
              Druskat, Stephan and Soares Siqueira, Abel and Garcia Gonzalez,
              Jesus and Cushing, Reggie},
    license = {Apache-2.0},
    month = aug,
    title = {{cffinit}},
    url = {https://github.com/citation-file-format/cff-initializer-javascript},
    version = {2.3.1},
    year = {2023},
}

@software{McArthur2025,
    author = {Sean McArthur},
    title = {{reqwest}: higher level HTTP client library},
    url = {https://github.com/seanmonstar/reqwest},
    date = {2025-10-13},
    version = {0.12.24},
    license = {MIT OR Apache-2.0},
}

@software{Haug2025,
    author = {Martin Haug},
    title = {{biblatex}: Parsing, writing, and evaluating BibTeX and BibLaTeX
             files},
    url = {https://github.com/typst/biblatex},
    date = {2025-09-25},
    version = {0.11.0},
    license = {MIT OR Apache-2.0},
}

@software{Saparelli2022,
    author = {Félix Saparelli},
    title = {{citeworks-cff}: Serde types for serialising and deserialising CFF
             (Citation File Format)},
    url = {https://github.com/passcod/citeworks},
    date = {2022-08-31},
    version = {0.1.1},
    license = {Apache-2.0},
}

\end{document}